# A Non-Triviality Certificate for Scalars and its application to Linear Systems

Deepak Ponvel Chermakani

deepakc@pmail.ntu.edu.sg  deepakc@e.ntu.edu.sg  deepakc@ed-alumni.net  deepakc@myfastmail.com  deepakc@usa.com

*Abstract: -* We present an approach of taking a linear weighted Average of *N* given scalars, such that this Average is zero, if and only if, all *N* scalars are zero. The weights for the scalars in this Average vary asymptotically with respect to a large positive real. We use this approach with a previous result on Asymptotic Linear Programming, to develop an *O(M^4)* Algorithm that decides whether or not a system of *M* Linear Inequalities is feasible, and, whether or not any desired subset of the variables in this system, is permitted to have a non-trivial solution.

## 1.   Introduction

Given a set of real (i.e. zero, positive or negative) scalars, we can decide whether or not the set is trivial (i.e. all scalars are zero) by comparing each scalar to zero. In this paper, we develop a triviality Certificate (i.e. a test to decide on triviality) for the scalars, which can be readily applied in a Linear System, consisting of strict and non-strict inequalities and equations. In our paper, if A and B are two Boolean statements, (A→B) denotes that (A is true implies B is true), and (A↔B) denotes that (A is true, if and only, if B is true). Next, if *a* and *b* are two scalars, *a*b* = *ab* = (product of *a* and *b*).

## 2.   The Certificate of Triviality for a given Set of Scalars

**Theorem-1**: There exists a positive real $\gamma$ that is a function of the given real scalars $\{x_1, x_2, \ldots x_N\}$, and there exists a real variable $K$ such that for all $K > \gamma$, the following statement is true:
$((x_i = 0 \text{ for all integers } i \text{ in } [1,N]) \leftrightarrow (((x_1/(K+1)) + (x_2/(K+2)) + \ldots + (x_N/(K+N))) = 0))$

**Proof:** It is obvious that $(x_i = 0 \text{ for all integers } i \text{ in } [1,N]) \rightarrow (((x_1/(K+1)) + (x_2/(K+2)) + \ldots + (x_N/(K+N))) = 0)$. Next, we focus on proving $(((x_1/(K+1)) + (x_2/(K+2)) + \ldots + (x_N/(K+N))) = 0) \rightarrow (x_i = 0 \text{ for all integers } i \text{ in } [1,N])$.

Expressing $((x_1/(K+1)) + (x_2/(K+2)) + \ldots + (x_N/(K+N)))$ as a single rational expression, we obtain:
$((x_1 A_1 + x_2 A_2 + \ldots + x_N A_N) / ((K+1)(K+2)\ldots(K+N)))$, where, for all integers $i$ in $[1, N]$, $A_i$ = (product of $(K+j)$, over all integers $j$ in $[1,N]$ and $j \neq i$). We can write the expression $(x_1 A_1 + x_2 A_2 + \ldots + x_N A_N)$ as $(K^{N-1} B_{N-1} + K^{N-2} B_{N-2} + \ldots + K^0 B_0)$, where, for all integers $i$ in $[0, N-1]$, $B_i$ represents the coefficient of $K^i$ in the expression $(x_1 A_1 + x_2 A_2 + \ldots + x_N A_N)$. We have:
$B_{N-1} = x_1 + x_2 + \ldots + x_N$
$B_{N-2} = (2+3+\ldots+N)x_1 + (1+3+4+\ldots N)x_2 + (1+2+4+5+\ldots+N)x_3 + \ldots + (1+2+3+\ldots+(N-1))x_N$
$B_{N-3} = (2*3+2*4+\ldots+2*N + 3*4 + 3*5 + \ldots 3*N + \ldots (N-1)*N)x_1 +$
  $(1*3 + 1*4 \ldots 1*N + 3*4 + 3*5 + \ldots 3*N + \ldots (N-1)*N)x_2 +$
  $\ldots +$
  $(1*2+1*3+\ldots+1*(N-1) + 2*3+2*4+\ldots+2*(N-1) + (N-2)*(N-1))x_N$
$\ldots$
$B_0 = (2*3*\ldots*N)x_1 + (1*3*4*\ldots*N)x_2 + (1*2*4*5*\ldots*N)x_3 + \ldots + (1*2*3*\ldots*(N-1))x_N$

Generalizing the pattern in the above coefficients, $B_{N-1} = x_1 + x_2 + \ldots + x_N$, and, for all integers $i$ in $[0,(N-2)]$, $B_i$ = (Summation over all integers $j$ in $[1,N]$, of $(x_j*$(summation of all combinations of product terms from Set of elements $\{\{1,2,\ldots N\} - \{j\}\}$, having $(N-i-1)$ elements in each product term$)))$.

Now consider the expression $(K^{N-1} B_{N-1} + K^{N-2} B_{N-2} + \ldots + K^0 B_0)$ as a univariate Polynomial in $K$. For a given set of scalars $\{x_1, x_2, \ldots x_N\}$, it is obvious that there exists an upper bound $\gamma$ on the real root of this Polynomial, given by Lagrange's Theorem [1]. Hence for all $K > \gamma$, the only possibility for $((K^{N-1} B_{N-1} + K^{N-2} B_{N-2} + \ldots + K^0 B_0) = 0)$ to be true, is $(B_i = 0$, for all integers $i$ in $[0, N-1])$. This gives us a set of $N$ linear equations in $\{x_1, x_2, \ldots x_N\}$, mentioned below in Lemma-1:

**Lemma-1**: The following $N$ linear equations in $\{x_1, x_2, \ldots x_N\}$ are unique:
$B_{N-1} = x_1 + x_2 + \ldots + x_N = 0$, and,
for all integers $i$ in $[0,(N-2)]$, $B_i$ = (Summation over all integers $j$ in $[1,N]$, of $(x_j*Comb_{(N-i-1)}(\{j\}))) = 0$.

Here $(Comb_{(N-i-1)}(\{j\}))$ denotes summation of all combinations of product terms from Set of elements $\{\{1,2,\ldots N\} - \{j\}\}$, having $(N-i-1)$ elements in each product term. We denote Set$\{a,b,c,d\}$ − Set$\{b,d\}$ = Set $\{a,c\}$.

Proof: (These $N$ linear equations are unique) ↔ (determinant of matrix $\Omega_l$, formed from coefficients of the linear equations, is non-zero). $\Omega_l$ is shown in the Figure 1. Also, (determinant of a matrix is zero) ↔ (determinant of its transpose is 0).

| 1 | 1 | ... | ... | 1 | 1 |
|---|---|---|---|---|---|
| $Comb_{(1)}(\{1\})$ | $Comb_{(1)}(\{2\})$ | ... | ... | $Comb_{(1)}(\{N-1\})$ | $Comb_{(1)}(\{N\})$ |
| $Comb_{(2)}(\{1\})$ | $Comb_{(2)}(\{2\})$ | ... | ... | $Comb_{(2)}(\{N-1\})$ | $Comb_{(2)}(\{N\})$ |
| ... | ... | ... | ... | ... | ... |
| ... | ... | ... | ... | ... | ... |
| $Comb_{(N-2)}(\{1\})$ | $Comb_{(N-2)}(\{2\})$ | ... | ... | $Comb_{(N-2)}(\{N-1\})$ | $Comb_{(N-2)}(\{N\})$ |
| $Comb_{(N-1)}(\{1\})$ | $Comb_{(N-1)}(\{2\})$ | ... | ... | $Comb_{(N-1)}(\{N-1\})$ | $Comb_{(N-1)}(\{N\})$ |

**Figure 1:** The square matrix $\Omega_1$ of dimension $N$

Denoting Column $i$ in the matrix as $C_i$, we apply column operations $C_{i\_next} = C_i - C_{i+1}$ on $\Omega_1$, for all integers $i$ in $[1, N-1]$. This eliminates one dimension, and we get the next square matrix $\Omega_2$ of dimension $N-1$, shown in Figure 2.

| 1 | 1 | ... | ... | 1 | 1 |
|---|---|---|---|---|---|
| $1*Comb_{(1)}(\{1,2\})$ | $1*Comb_{(1)}(\{2,3\})$ | ... | ... | $1*Comb_{(1)}(\{N-2,N-1\})$ | $1*Comb_{(1)}(\{N-1,N\})$ |
| $1*Comb_{(2)}(\{1,2\})$ | $1*Comb_{(2)}(\{2,3\})$ | ... | ... | $1*Comb_{(2)}(\{N-2,N-1\})$ | $1*Comb_{(2)}(\{N-1,N\})$ |
| ... | ... | ... | ... | ... | ... |
| ... | ... | ... | ... | ... | ... |
| $1*Comb_{(N-3)}(\{1,2\})$ | $1*Comb_{(N-3)}(\{2,3\})$ | ... | ... | $1*Comb_{(N-3)}(\{N-2,N-1\})$ | $1*Comb_{(N-3)}(\{N-1,N\})$ |
| $1*Comb_{(N-2)}(\{1,2\})$ | $1*Comb_{(N-2)}(\{2,3\})$ | ... | ... | $1*Comb_{(N-2)}(\{N-2,N-1\})$ | $1*Comb_{(N-2)}(\{N-1,N\})$ |

**Figure 2:** The square matrix $\Omega_2$ of dimension $N-1$

Again apply $C_{i\_next} = C_i - C_{i+1}$ for all integers $i$ in $[1, N-2]$, to eliminate another dimension to get square matrix $\Omega_3$ in Figure 3.

| 2 | 2 | ... | 2 | 2 |
|---|---|---|---|---|
| $2*Comb_{(1)}(\{1,2,3\})$ | $2*Comb_{(1)}(\{2,3,4\})$ | ... | $2*Comb_{(1)}(\{N-3,N-2,N-1\})$ | $2*Comb_{(1)}(\{N-2,N-1,N\})$ |
| $2*Comb_{(2)}(\{1,2,3\})$ | $2*Comb_{(2)}(\{2,3,4\})$ | ... | $2*Comb_{(2)}(\{N-3,N-2,N-1\})$ | $2*Comb_{(2)}(\{N-2,N-1,N\})$ |
| ... | ... | ... | ... | ... |
| $2*Comb_{(N-4)}(\{1,2,3\})$ | $2*Comb_{(N-4)}(\{2,3,4\})$ | ... | $2*Comb_{(N-4)}(\{N-3,N-2,N-1\})$ | $2*Comb_{(N-4)}(\{N-2,N-1,N\})$ |
| $2*Comb_{(N-3)}(\{1,2,3\})$ | $2*Comb_{(N-3)}(\{2,3,4\})$ | ... | $2*Comb_{(N-3)}(\{N-3,N-2,N-1\})$ | $2*Comb_{(N-3)}(\{N-2,N-1,N\})$ |

**Figure 3:** The square matrix $\Omega_3$ of dimension $N-2$

Divide all columns of $\Omega_3$ by $2$, and again apply column operations $C_{i\_next} = C_i - C_{i+1}$, for all integers $i$ in $[1, N-3]$. This eliminates another dimension, and we get square matrix $\Omega_4$ of dimension $N-3$, in Figure 4.

| 3 | 3 | ... | 3 | 3 |
|---|---|---|---|---|
| $3*Comb_{(1)}(\{1,2,3,4\})$ | $3*Comb_{(1)}(\{2,3,4,5\})$ | ... | $3*Comb_{(1)}(\{N-4,N-3,N-2,N-1\})$ | $3*Comb_{(1)}(\{N-3,N-2,N-1,N\})$ |
| $3*Comb_{(2)}(\{1,2,3,4\})$ | $3*Comb_{(2)}(\{2,3,4,5\})$ | ... | $3*Comb_{(2)}(\{N-4,N-3,N-2,N-1\})$ | $3*Comb_{(2)}(\{N-3,N-2,N-1,N\})$ |
| ... | ... | ... | ... | ... |
| $3*Comb_{(N-5)}(\{1,2,3,4\})$ | $3*Comb_{(N-5)}(\{2,3,4,5\})$ | ... | $3*Comb_{(N-5)}(\{N-4,N-3,N-2,N-1\})$ | $3*Comb_{(N-5)}(\{N-3,N-2,N-1,N\})$ |
| $3*Comb_{(N-4)}(\{1,2,3,4\})$ | $3*Comb_{(N-4)}(\{2,3,4,5\})$ | ... | $3*Comb_{(N-4)}(\{N-4,N-3,N-2,N-1\})$ | $3*Comb_{(N-4)}(\{N-3,N-2,N-1,N\})$ |

**Figure 4:** The square matrix $\Omega_4$ of dimension $N-3$

We will now show by Induction, that dividing all columns of $\Omega_j$ by $(j-1)$, where $2 \leq j \leq (N-1)$, and subsequently applying column operations $C_{i\_next} = C_i - C_{i+1}$, for all integers $i$ in $[1, N-j]$, we get square matrix $\Omega_{j+1}$ of dimension $(N-j)$, in Figure 5:

| j | j | ... | j | j |
|---|---|---|---|---|
| $j*Comb_{(1)}(\{1,2,...,j+1\})$ | $j*Comb_{(1)}(\{2,3,...,j+2\})$ | ... | $j*Comb_{(1)}(\{N-j-1,...,N-2,N-1\})$ | $j*Comb_{(1)}(\{N-j,...,N-1,N\})$ |
| $j*Comb_{(2)}(\{1,2,...,j+1\})$ | $j*Comb_{(2)}(\{2,3,...,j+2\})$ | ... | $j*Comb_{(2)}(\{N-j-1,...,N-2,N-1\})$ | $j*Comb_{(2)}(\{N-j,...,N-1,N\})$ |
| ... | ... | ... | ... | ... |
| $j*Comb_{(N-j-2)}(\{1,2,...,j+1\})$ | $j*Comb_{(N-j-2)}(\{2,3,...,j+2\})$ | ... | $j*Comb_{(N-j-2)}(\{N-j-1,...,N-2,N-1\})$ | $j*Comb_{(N-j-2)}(\{N-j,...,N-1,N\})$ |
| $j*Comb_{(N-j-1)}(\{1,2,...,j+1\})$ | $j*Comb_{(N-j-1)}(\{2,3,...,j+2\})$ | ... | $j*Comb_{(N-j-1)}(\{N-j-1,...,N-2,N-1\})$ | $j*Comb_{(N-j-1)}(\{N-j,...,N-1,N\})$ |

**Figure 5:** The square matrix $\Omega_{j+1}$ of dimension $(N-j)$

Consider any column vector $C_i$ in $\Omega_j$ ($1 \leq i \leq (N-j)$). Assume that the first element in $C_i$ is $(j+1)$, and the $k^{th}$ element ($2 \leq k \leq (N-j+1)$) in $C_i$ is $((j+1)*Comb_{(k-1)} (\{i,i+1,...,i+j-1\}))$. Further assume that the first element in $C_{i+1}$ is $(j+1)$, and the $k^{th}$ element ($2 \leq k \leq (N-j+1)$) in $C_{i+1}$ is $((j+1)*Comb_{(k-1)} (\{i+1,i+2,...,i+j\}))$. After dividing all elements of $\Omega_j$ by $(j+1)$, the value of the $k^{th}$ element in $(C_i - C_{i+1})$ becomes:

$((Comb_{(k-1)} (\{i,i+1,...,i+j-1\})) - (Comb_{(k-1)} (\{i+1,i+2,...,i+j\})))$, which is equal to
$((i+j)*Comb_{(k-2)} (\{i,i+1,...,i+j\}) - (i)*Comb_{(k-2)} (\{i,i+1,...,i+j\}))$, which is equal to
$(j* Comb_{(k-2)} (\{i,i+1,...,i+j\}))$.

The loss of dimension is obvious after applying $C_{i\_next} = (C_i - C_{i+1})$, for all integers $i$ in $[1,N-j]$, since the first row of $\Omega_j$ always has 1, after the division of all elements of $\Omega_j$ by $(j+1)$. $\Omega_{j+1}$ of Figure 5 is thus proved to be obtained by Induction.

We proceed to iteratively obtain square matrices of smaller dimensions, until $\Omega_{N-1}$ of dimension 2 in Figure 6.

| N-2 | N-2 |
|---|---|
| (N-2)*N | N-2 |

Figure 6: The square matrix $\Omega_{N-1}$ of dimension 2

The final operation of dividing all columns of $\Omega_{N-1}$ by $(N-2)$ and applying the column operation $C_{1\_next} = C_1 - C_2$, yields the single element $(N-1)$, which is non-zero for all $N>1$. **Hence proved Lemma-1**.

Thus, the only solution that satisfies the set of homogenous linear equations in Lemma-1, is $x_i = 0$ for all integers $i$ in $[1,N]$. **Hence proved Theorem-1**.

Start of example illustrating Theorem-1
As an example with $N=5$, the expression: $((x_1/(K+1)) + (x_2/(K+2)) + (x_3/(K+3)) + (x_4/(K+4)) + (x_5/(K+5)))$
$= ((K+2)(K+3)(K+4)(K+5)x_1 + (K+1)(K+3)(K+4)(K+5)x_2 + (K+1)(K+2)(K+4)(K+5)x_3 + (K+1)(K+2)(K+3)(K+5)x_4 + (K+1)(K+2)(K+3)(K+4)x_5) / (K+1)(K+2)(K+3)(K+4)(K+5)$
$= ($
$K^4(x_1 + x_2 + x_3 + x_4 + x_5) + K^3((2+3+4+5)x_1 + (1+3+4+5)x_2 + (1+2+4+5)x_3 + (1+2+3+5)x_4 + (1+2+3+4)x_5) +$
$K^2((2*3+2*4+2*5+3*4+3*5+4*5)x_1 + (1*3+1*4+1*5+3*4+3*5+4*5)x_2 + (1*2+1*4+1*5+2*4+2*5+4*5)x_3 +$
$(1*2+1*3+1*5+2*3+2*5+3*5)x_4 + (1*2+1*3+1*4+2*3+2*4+3*4)x_5) +$
$K ((2*3*4+2*3*5+2*4*5+3*4*5)x_1 + (1*3*4+1*3*5+1*4*5+3*4*5)x_2 + (1*2*4+1*2*5+1*4*5+2*4*5)x_3 +$
$(1*2*3+1*2*5+1*3*5+2*3*5)x_4 + (1*2*3+1*2*4+1*3*4+2*3*4)x_5) +$
$((2*3*4*5)x_1 + (1*3*4*5)x_2 + (1*2*4*5)x_3 + (1*2*3*5)x_4 + (1*2*3*4)x_5)$
$) / (K+1)(K+2)(K+3)(K+4)(K+5)$

The matrix $\Omega_1$ is shown in Figure 7.

| 1 | 1 | 1 | 1 | 1 |
|---|---|---|---|---|
| 2+3+4+5 | 1+3+4+5 | 1+2+4+5 | 1+2+3+5 | 1+2+3+4 |
| 2*3+2*4+2*5+3*4+3*5+4*5 | 1*3+1*4+1*5+3*4+3*5+4*5 | 1*2+1*4+1*5+2*4+2*5+4*5 | 1*2+1*3+1*5+2*3+2*5+3*5 | 1*2+1*3+1*4+2*3+2*4+3*4 |
| 2*3*4+2*3*5+2*4*5+3*4*5 | 1*3*4+1*3*5+1*4*5+3*4*5 | 1*2*4+1*2*5+1*4*5+2*4*5 | 1*2*3+1*2*5+1*3*5+2*3*5 | 1*2*3+1*2*4+1*3*4+2*3*4 |
| 2*3*4*5 | 1*3*4*5 | 1*2*4*5 | 1*2*3*5 | 1*2*3*4 |

Figure 7: The square matrix $\Omega_1$ for our example

Applying $C_{1\_next} = C_1 - C_2$, $C_{2\_next} = C_2 - C_3$, $C_{3\_next} = C_3 - C_4$, $C_{4\_next} = C_4 - C_5$, to $\Omega_1$, we get rid of the first row and last column, and the resulting matrix $\Omega_2$ is shown in Figure 8.

| 1 | 1 | 1 | 1 |
|---|---|---|---|
| 3+4+5 | 1+4+5 | 1+2+5 | 1+2+3 |
| 3*4+3*5+4*5 | 1*4+1*5+4*5 | 1*2+1*5+2*5 | 1*2+1*3+2*3 |
| 3*4*5 | 1*4*5 | 1*2*5 | 1*2*3 |

Figure 8: The square matrix $\Omega_2$ for our example

Applying $C_{1\_next} = C_1 - C_2$, $C_{2\_next} = C_2 - C_3$, $C_{3\_next} = C_3 - C_4$, to $\Omega_2$, we get rid of the first row and last column, and the resulting matrix $\Omega_3$ is shown in Figure 9.

|   2    |   2    |   2    |
|--------|--------|--------|
| 2*(4+5)| 2*(1+5)| 2*(1+2)|
| 2*(4*5)| 2*(1*5)| 2*(1*2)|

**Figure 9: The square matrix $\Omega_3$ for our example**

Dividing all elements of $\Omega_3$ by 2, and then applying $C_{1\_next} = C_1 - C_2$, $C_{2\_next} = C_2 - C_3$, to $\Omega_3$, we get rid of the first row and last column, and the resulting matrix $\Omega_4$ is shown in Figure 10.

|   3   |   3   |
|-------|-------|
|  3*5  |  3*1  |

**Figure 10: The square matrix $\Omega_4$ for our example**

Dividing all elements of $\Omega_4$ by 3, and then applying $C_{1\_next} = C_1 - C_2$, to $\Omega_4$ yields the single element of $5 - 1 = 4$.
End of Example illustrating Theorem-1

## 3. Application to deciding existence of non-trivial feasible solutions to Linear Systems

In this section, we show how Theorem-1 can be used to decide whether or not non-trivial feasible solutions can exist to Linear Systems. The Linear System we will consider is a set of linear constraints over continuous real variables (i.e. the variables are allowed to take the values of zero, positive Reals, or negative Reals), which we shall refer to as $S_{linear}$, having $P$ non-strict linear inequalities and $Q$ strict linear inequalities shown below:

$a_{1,1} y_1 + a_{1,2} y_2 + ... + a_{1,N} y_N \leq c_1$
$a_{2,1} y_1 + a_{2,2} y_2 + ... + a_{2,N} y_N \leq c_2$
...
$a_{P,1} y_1 + a_{P,2} y_2 + .. + a_{P,N} y_N \leq c_P$
$b_{1,1} y_1 + b_{1,2} y_2 + ... + b_{1,N} y_N < d_1$
$b_{2,1} y_1 + b_{2,2} y_2 + ... + b_{2,N} y_N < d_2$
...
$b_{Q,1} y_1 + b_{Q,2} y_2 + .. + b_{Q,N} y_N < d_Q$

In $S_{linear}$, for all integers $j$ in $[1,N]$, for all integers $i$ in $[1,P]$, for all integers $k$ in $[1,Q]$: $y_j$ is a real variable, and the elements of $\{a_{i,j}, c_i, b_{k,j}, d_k\}$ belong to the set of integers.

It is well-known that a linear equality $(x = a)$ can be expressed as a set of two non-strict linear inequalities $((x \leq a)$ AND $(x \geq a))$. Hence, $S_{linear}$ is able to express most linear systems, except of course linear discrete systems (for example, if $x$ is constrained to integers or binary values).

The authors [2][3][4] have described methods for deciding whether or not $S_{linear}$ is feasible, and if so, then finding the non-trivial solutions of $S_{linear}$, if they exist. In this paper, we give additional focus to deciding whether or not a feasible solution is permitted to $S_{linear}$, where any desired subset of the variables are non-trivial (for example, can the subset $\{y_2, y_5, y_{13}, y_N\}$ be non-trivial in a feasible solution to $S_{linear}$?). This question can be of potential use in real-life engineering problems, where a subset of the universal set of variables $\{y_1, y_2, ... y_N\}$, is going to be applicable in a subsequent context.

So we define our problem $P_{linear}$ to be consisting of two parts:
1) Decide whether or not $S_{linear}$ is feasible, and,
2) If the answer to part-1 is YES, decide whether or not a feasible solution is permitted to $S_{linear}$, where any desired subset of the variables is non-trivial.

We approach $P_{linear}$ by introducing an extra constraint, based on Theorem-1 of this paper, assuming non-triviality of the desired subset of variables. Since, this constraint involves setting a weighted average of the variables to be not equal to zero, the subsequent steps need to be repeated twice, once assuming that the *(weighted average is > 0)*, and once again assuming that the *(weighted average < 0)*.

We give the detailed steps of our Algorithm for $P_{linear}$ as follows, assuming without loss of generality, that we desire to determine whether or not the subset $\{y_2, y_5, y_{13}, y_N\}$ is allowed to be non-trivial in a feasible solution of $S_{linear}$.

Step-1: Convert $S_{linear}$ to a new system $S_{nonstrict\_linear}$ consisting entirely of non-strict linear inequalities, where the coefficients of the variables vary linearly with respect to a parameter $K$ referred to as the *time* parameter that tends to positive infinity. Note that all strict inequalities of the form $(x < a)$ can be converted to a set of non-strict inequalities of the form $(((a–x) \geq e)$ AND $((K*e) \geq 1))$, where $e$ is the extra variable introduced.

Step-2: Use existing Algorithms [5][6] on Asymptotic Linear Programming of complexity $O(M^4)$ for determining whether or not $S_{nonstrict\_linear}$ admits a feasible solution. Here, $M$ is the number of constraints in $S_{nonstrict\_linear}$, which is bounded by a linear multiple of the number of constraints in $S_{linear}$. If a feasible solution is allowed, then the answer to part-1 of $P_{linear}$ is YES and proceed to Step-3, else the answer to part-1 of $P_{linear}$ is NO and exit Algorithm.

Step-3: Form a new system of linear constraints called $S_{linear\_1}$, where $S_{linear\_1}$ is the union of $S_{linear}$ AND the constraint: $(((y_2/(K+1)) + (y_5/(K+2)) + (y_{13}/(K+3)) + (y_N/(K+4))) > 0)$.

Step-4: Convert $S_{linear\_1}$ to a new system $S_{nonstrict\_linear\_1}$ consisting entirely of non-strict linear inequalities, where the coefficients of the variables vary linearly with respect to the *time* parameter $K$. This can be achieved by following two sub-procedures. First, substitute all $(y_j/(K+k))$ with a new variable. For example, substitute $y_{13}/(K+3)$ with $z_{13}$, so that $y_{13} = (K+3)z_{13}$. Second, convert all strict inequalities to a set of non-strict inequalities as described in Step-1.

Step-5: Use existing Algorithms [5][6] on Asymptotic Linear Programming of complexity $O(M^4)$ for determining whether or not $S_{nonstrict\_linear\_1}$ admits a feasible solution.

Step-6: Form a new system of linear constraints called $S_{linear\_2}$, where $S_{linear\_2}$ is the union of $S_{linear}$ AND the constraint: $(((y_2/(K+1)) + (y_5/(K+2)) + (y_{13}/(K+3)) + (y_N/(K+4))) < 0)$.

Step-7: Convert $S_{linear\_2}$ to a new system $S_{nonstrict\_linear\_2}$ consisting entirely of non-strict linear inequalities, where the coefficients of the variables vary linearly with respect to the *time* parameter $K$, using the two sub-procedures in Step-4.

Step-8: Use existing Algorithms [5][6] on Asymptotic Linear Programming of complexity $O(M^4)$ for determining whether or not $S_{nonstrict\_linear\_2}$ admits a feasible solution.

Step-9: If either $S_{nonstrict\_linear\_1}$ or $S_{nonstrict\_linear\_2}$ admit a feasible solution, then the answer to part-2 of $P_{linear}$ is YES, else the answer to part-2 of $P_{linear}$ is NO.

A Note on Aymptotic Linear Programming
An Asymptotic Linear Program [5][6][7] is a linear program where the coefficients of the variables in the constraints are rational Polynomials involving a single parameter referred to as the *time* parameter. The author of [7] proved that as this *time* parameter grows beyond a certain positive value, the Linear Program gets constant properties such as feasibility/infeasibility, boundedness, consistency, and bounded constraint sets. In other words, as this *time* parameter tends to infinity, the Asymptotic Linear Program shows a stable steady-state behavior. The subsequent algorithms [5][6] developed to determine feasibility of Asymptotic Linear Programs, take advantage of this fact.

To make our Algorithm clearer, we give the following example:

Start of example illustrating the Algorithm for $P_{linear}$

Consider $S_{linear}$ to be defined by the following 4 constraints:
$$a_{1,1} y_1 + a_{1,2} y_2 + a_{1,3} y_3 \leq c_1$$
$$a_{2,1} y_1 + a_{2,2} y_2 + a_{2,3} y_3 \leq c_2$$
$$b_{1,1} y_1 + b_{1,2} y_2 + b_{1,3} y_3 < d_1$$
$$b_{2,1} y_1 + b_{2,2} y_2 + b_{2,3} y_3 < d_2$$

Suppose we have to decide whether or not the subset of variables $\{y_2, y_3\}$ can be non-trivial in a feasible solution of $S_{linear}$.

In Step-1, $S_{nonstrict\_linear}$ becomes as follows:
$$a_{1,1} y_1 + a_{1,2} y_2 + a_{1,3} y_3 \leq c_1$$
$$a_{2,1} y_1 + a_{2,2} y_2 + a_{2,3} y_3 \leq c_2$$
$$d_1 - b_{1,1} y_1 - b_{1,2} y_2 - b_{1,3} y_3 \geq e$$
$$d_2 - b_{2,1} y_1 - b_{2,2} y_2 - b_{2,3} y_3 \geq e$$
$$K*e \geq 1$$

In Step-2, if $S_{nonstrict\_linear}$ permits a feasible solution, part-1 of $P_{linear}$ is YES so go to Step-3, else part-1 of $P_{linear}$ is NO so exit.

In Step-3, $S_{linear\_1}$ becomes as follows:
$$a_{1,1} y_1 + a_{1,2} y_2 + a_{1,3} y_3 \leq c_1$$
$$a_{2,1} y_1 + a_{2,2} y_2 + a_{2,3} y_3 \leq c_2$$
$$b_{1,1} y_1 + b_{1,2} y_2 + b_{1,3} y_3 < d_1$$
$$b_{2,1} y_1 + b_{2,2} y_2 + b_{2,3} y_3 < d_2$$
$$(y_2/(K+1)) + (y_3/(K+2)) > 0$$

In Step-4, $S_{nonstrict\_linear\_1}$ becomes as follows:
$$a_{1,1} z_1 + (K+1)a_{1,2} z_2 + (K+2)a_{1,3} z_3 \leq c_1$$
$$a_{2,1} z_1 + (K+1)a_{2,2} z_2 + (K+2)a_{2,3} z_3 \leq c_2$$
$$d_1 - b_{1,1} z_1 - (K+1)b_{1,2} z_2 - (K+2)b_{1,3} z_3 \geq e$$
$$d_2 - b_{2,1} z_1 - (K+1)b_{2,2} z_2 - (K+2)b_{2,3} z_3 \geq e$$
$$z_2 + z_3 \geq e$$

$K*e \geq 1$

In Step-5, determine whether or not $S_{nonstrict\_linear\_1}$ admits a feasible solution.

In Step-6, $S_{linear\_2}$ becomes as follows:

$a_{1,1} y_1 + a_{1,2} y_2 + a_{1,3} y_3 \leq c_1$
$a_{2,1} y_1 + a_{2,2} y_2 + a_{2,3} y_3 \leq c_2$
$b_{1,1} y_1 + b_{1,2} y_2 + b_{1,3} y_3 < d_1$
$b_{2,1} y_1 + b_{2,2} y_2 + b_{2,3} y_3 < d_2$
$(y_2 / (K+1)) + (y_3 / (K+2)) < 0$

In Step-7, $S_{nonstrict\_linear\_2}$ becomes as follows:

$a_{1,1} z_1 + (K+1)a_{1,2} z_2 + (K+2)a_{1,3} z_3 \leq c_1$
$a_{2,1} z_1 + (K+1)a_{2,2} z_2 + (K+2)a_{2,3} z_3 \leq c_2$
$d_1 - b_{1,1} z_1 - (K+1)b_{1,2} z_2 - (K+2)b_{1,3} z_3 \geq e$
$d_2 - b_{2,1} z_1 - (K+1)b_{2,2} z_2 - (K+2)b_{2,3} z_3 \geq e$
$z_2 + z_3 \leq -e$
$K*e \geq 1$

In Step-8, determine whether or not $S_{nonstrict\_linear\_2}$ admits a feasible solution.

In Step-9, if either $S_{nonstrict\_linear\_1}$ or $S_{nonstrict\_linear\_2}$ admits a feasible solution, the answer to part-2 of $P_{linear}$ is YES, else the answer to part-2 of $P_{linear}$ is NO. Algorithm ends.

End of example illustrating the Algorithm for $P_{linear}$

## 4. Conclusion

In this paper, we presented an approach to obtain a linear weighted Average of $N$ given scalars, such that the value of this average is zero, if and only if, all the scalars are zero. This approach can be used to decide whether or not a desired subset of variables in a Linear System is allowed to be non-trivial in a feasible solution of the Linear System. The Linear System considered contains both strict and non-strict inequalities. After plugging in the approach into the given linear system, an equivalent set of Asymptotic Linear Constraints with only non-strict inequalities is generated, such that the new Asymptotic Linear System has a feasible solution, if and only if, the given Linear System has a feasible solution. Existing algorithms for Asymptotic Linear Programming are finally used to decide feasibility of the resulting Asymptotic Linear System.

## 5. Future Work

Though we have considered Linear Systems with linear strict and non-strict inequalities, we are unable to efficiently consider linear Inequations (i.e. constraints of the form $x \neq a$). Currently, the only possible approach (and which has been applied in the Algorithm of this paper) seems to be to repeat subsequent Algorithm steps twice (i.e. once with $x < a$, and again with $x > a$). If there were to be multiple such Inequations in a Linear System, we would need computational effort that grows exponentially with the number of such Inequations. Hence, more work needs to be done to efficiently model Inequations.

**About the Author**
I, Deepak Ponvel Chermakani, wrote this paper, out of my own interest and initiative, during my spare time. In Sep-2010, I completed a fulltime one year Master Degree course in *Operations Research with Computational Optimization* from University of Edinburgh UK (www.ed.ac.uk). In Jul-2003, I completed a fulltime four year Bachelor Degree course in *Electrical and Electronic Engineering*, from Nanyang Technological University Singapore (www.ntu.edu.sg). I completed my fulltime high schooling from National Public School in Bangalore in India in Jul-1999.